\documentclass[aps,prd,superscriptaddress,onecolumn,floatfix,nofootinbib,preprintnumbers]{revtex4}

\usepackage{amsmath,graphicx,tabularx,url}

\begin{document}

\def\tablename{Table}
\def\figurename{Figure}

\def\gtot{\Gamma_\text{tot}}
\def\brinv{\text{BR}_\text{inv}}
\def\brsm{\text{BR}_\text{SM}}
\def\bratio{\mathcal{B}_\text{inv}}
\def\as{\alpha_s}
\def\az{\alpha_0}
\def\gz{g_0}
\def\w{\vec{w}}
\def\sdag{\Sigma^{\dag}}
\def\s{\Sigma}
\newcommand{\psib}{\overline{\psi}}
\newcommand{\Psib}{\overline{\Psi}}
\newcommand\one{\leavevmode\hbox{\small1\normalsize\kern-.33em1}}
\newcommand{\Mpl}{M_\mathrm{Pl}}
\newcommand{\p}{\partial}
\newcommand{\lag}{\mathcal{L}}
\newcommand{\qqquad}{\qquad \qquad}
\newcommand{\qqqquad}{\qquad \qquad \qquad}

\newcommand{\qb}{\bar{q}}
\newcommand{\matx}{|\mathcal{M}|^2}
\newcommand{\really}{\stackrel{!}{=}}
\newcommand{\msbar}{\overline{\text{MS}}}
\newcommand{\qns}{f_q^\text{NS}}
\newcommand{\lqcd}{\Lambda_\text{QCD}}
\newcommand{\met}{\slashchar{p}_T}
\newcommand{\ptmiss}{\slashchar{\vec{p}}_T}
\newcommand{\pmiss}{\slashchar{p}}

\newcommand{\st}[1]{\tilde{t}_{#1}}
\newcommand{\stb}[1]{\tilde{t}_{#1}^*}
\newcommand{\nz}[1]{\tilde{\chi}_{#1}^0}
\newcommand{\cp}[1]{\tilde{\chi}_{#1}^+}
\newcommand{\cm}[1]{\tilde{\chi}_{#1}^-}

\providecommand{\mg}{m_{\tilde{g}}}
\providecommand{\mst}{m_{\tilde{t}}}
\newcommand{\msn}[1]{m_{\tilde{\nu}_{#1}}}
\newcommand{\mch}[1]{m_{\tilde{\chi}^+_{#1}}}
\newcommand{\mne}[1]{m_{\tilde{\chi}^0_{#1}}}
\newcommand{\msb}[1]{m_{\tilde{b}_{#1}}}

\newcommand{\mev}{{\ensuremath\rm MeV}}
\newcommand{\gev}{{\ensuremath\rm GeV}}
\newcommand{\tev}{{\ensuremath\rm TeV}}
\newcommand{\fb}{{\ensuremath\rm fb}}
\newcommand{\ab}{{\ensuremath\rm ab}}
\newcommand{\pb}{{\ensuremath\rm pb}}
\newcommand{\sign}{{\ensuremath\rm sign}}
\newcommand{\ifb}{{\ensuremath\rm fb^{-1}}}

\def\slashchar#1{\setbox0=\hbox{$#1$}           
   \dimen0=\wd0                                 
   \setbox1=\hbox{/} \dimen1=\wd1               
   \ifdim\dimen0>\dimen1                        
      \rlap{\hbox to \dimen0{\hfil/\hfil}}      
      #1                                        
   \else                                        
      \rlap{\hbox to \dimen1{\hfil$#1$\hfil}}   
      /                                         
   \fi}
\newcommand{\dslash}{\slashchar{\partial}}
\newcommand{\Dslash}{\slashchar{D}}

\def\eg{{\sl e.g.} \,}
\def\ie{{\sl i.e.} \,}
\def\etal{{\sl et al} \,}

\title{Stop searches in 2012}

\author{Tilman Plehn}
\affiliation{Institut f\"ur Theoretische Physik, Universit\"at Heidelberg, Germany}

\author{Michael Spannowsky}
\affiliation{IPPP, Department of Physics, Durham University, United Kingdom}

\author{Michihisa Takeuchi}
\affiliation{Institut f\"ur Theoretische Physik, Universit\"at Heidelberg, Germany}

\begin{abstract}
For this year's 8~TeV run of the LHC we lay out different strategies
to search for scalar top pairs.  We show results for the hadronic and
for the semi-leptonic channels based on hadronic top tagging. For the
di-lepton channel we illustrate the impact of transverse mass
variables.  Each of our signal-to-background ratios ranges around
unity for a stop mass around 400~GeV. The combined signal
significances show that dedicated stop searches are becoming sensitive
over a non-negligible part of parameter space.
\end{abstract}

\preprint{IPPP/12/30}
\preprint{DCPT/12/60}

\maketitle

\tableofcontents 

\newpage

\section{Introduction}
\label{sec:intro}

The primary goal of the LHC is to unveil the mechanism which breaks
electroweak symmetry. The minimal solution, realized in the Standard
Model (SM), predicts the existence of a fundamental scalar field, the
Higgs boson. The perturbative instability of scalar masses is one of
the motivations for the existence of new physics at the electroweak
scale~\cite{bsm_review}. One of the proposed solutions to this problem
is supersymmetry at the TeV scale~\cite{primer}.\medskip

The leading contribution to the quadratic divergence of the scalar
Higgs mass in the Standard Model arises from the top quark with its
large Yukawa coupling. Its supersymmetric partner is the scalar top
quark. It can be pair produced at hadron colliders just through its
QCD couplings, \ie without any mediation by additional
supersymmetric particles. Its production cross section at the LHC
therefore only depends on the stop mass~\cite{prospino_stops}. The
decay channels of the stop depend almost entirely on the
supersymmetric mass spectrum~\cite{stop_decays,sdecay}: if
kinematically allowed, it will decay through its strong coupling into
a top quark and a gluino.  Below this threshold there exist two weak
decay channels, a charged current decay to bottom-chargino and a
neutral current decay
\begin{equation}
\st{1} \to t \nz{1} \; ,
\label{eq:decay}
\end{equation}
where we assume the lightest neutralino to be the dark matter agent
and hence appear as missing transverse momentum at hadron
colliders. This neutral current decay we target in this analysis.
For even smaller stop masses a loop-induced decay to a
charm quark and the lightest neutralino dominates.\medskip

At the Tevatron searches for such loop-induced decays as well as the
bottom-chargino signature have lead to moderate limits on the stop
mass~\cite{stops_tevatron_c,stops_tevatron_b}.  In 2011 the LHC
delivered approximately $5~\ifb$ of integrated luminosity to ATLAS and
CMS. This data set allowed both collaborations to significantly
constrain the squark-gluino mass plane through the search for jets
plus missing energy~\cite{ATLAS_susy,CMS_susy}.~\footnote{Note that
  replacing the squark-gluino mass plane for example by an
  $m_0-m_{1/2}$ plane in the CMSSM implies significant unwanted and
  unnecessary model assumptions~\cite{wedges}.}  Direct stop searches
at the LHC are notorious, mostly because of the small
signal cross section and the overwhelming background from top pair
production~\cite{HEP1}. Therefore, all limits on top partner searches
from 2011 analyses are derived from supersymmetric model
assumptions~\cite{natural}: either, the gluino has a sizable
branching ratio into 3rd generation squarks~\cite{gluino3rd_atlas} or
sbottom limits get re-interpreted in terms of the stop mass based on
the SU(2) symmetry of their left-handed modes.
More specifically, assuming gauge mediated supersymmetry breaking with
the NLSP decay $\nz1 \to Z\tilde{G}$ we can exclude masses below
$m_{\st1}<330$~GeV for $\mne{1}=190$~\cite{stop_gm_atlas}.\medskip

For 14~TeV collider energy we have shown that with the help of top
taggers~\cite{toptagger,tagger_review} we can reconstruct top quarks
from stop decays and extract stop pair production from top
backgrounds. This holds for purely hadronic decays of the two top
quarks from the stop pair~\cite{HEP1,HEP3} as well for the
semi-leptonic channel~\cite{HEP2}. The main benefits of the new top
reconstruction methods in these analyses is that they automatically
resolve any combinatorics of the top decay jets and fully reconstruct
the top 4-momenta and angular correlations~\cite{meade_reece} The
associated analysis is not any more complicated than a search for
slepton or sbottom pairs, including the application of $m_{T2}$ for a
stop mass measurement~\cite{mt2}.

In this paper we test a variety of search strategies for light stops
($350 \leq m_{\st{1}} \leq 700$ GeV) at 8~TeV. We see that a
straightforward adaption of the 14~TeV search strategies is
challenging, due to the small signal rate.  Instead, we can optimize
the signal efficiency by looser requirements on the reconstructed stop
decays while benefiting from the also significantly reduced top pair
background. This change with respect to the 14~TeV
analyses~\cite{HEP1,HEP2,HEP3} reduces the potential of reconstructing
masses and model parameters, but it will allow us to probe a
significant range of stop masses in 2012~\cite{Bai:2012gs}.\medskip

We organize this brief overview in the following way: In
Sec.~\ref{sec:event} we briefly review the properties of the stop
signal and of the SM backgrounds at 8~TeV. This includes production
cross sections and $p_T$ distributions.  In Sec.~\ref{sec:reconst} we
discuss our results for hadronic, semi-leptonic and leptonic top
decays. In the latter case we find that a simple transverse mass
variable removes almost the entire background. For each of these
channels we quote signal-to-background ratios and signal
significances.  A summary follows in Sec.~\ref{sec:conclusion}.

\newpage

\section{Stop production at 8~TeV}
\label{sec:event}

\begin{figure}[b]
\includegraphics[height=0.35\textwidth]{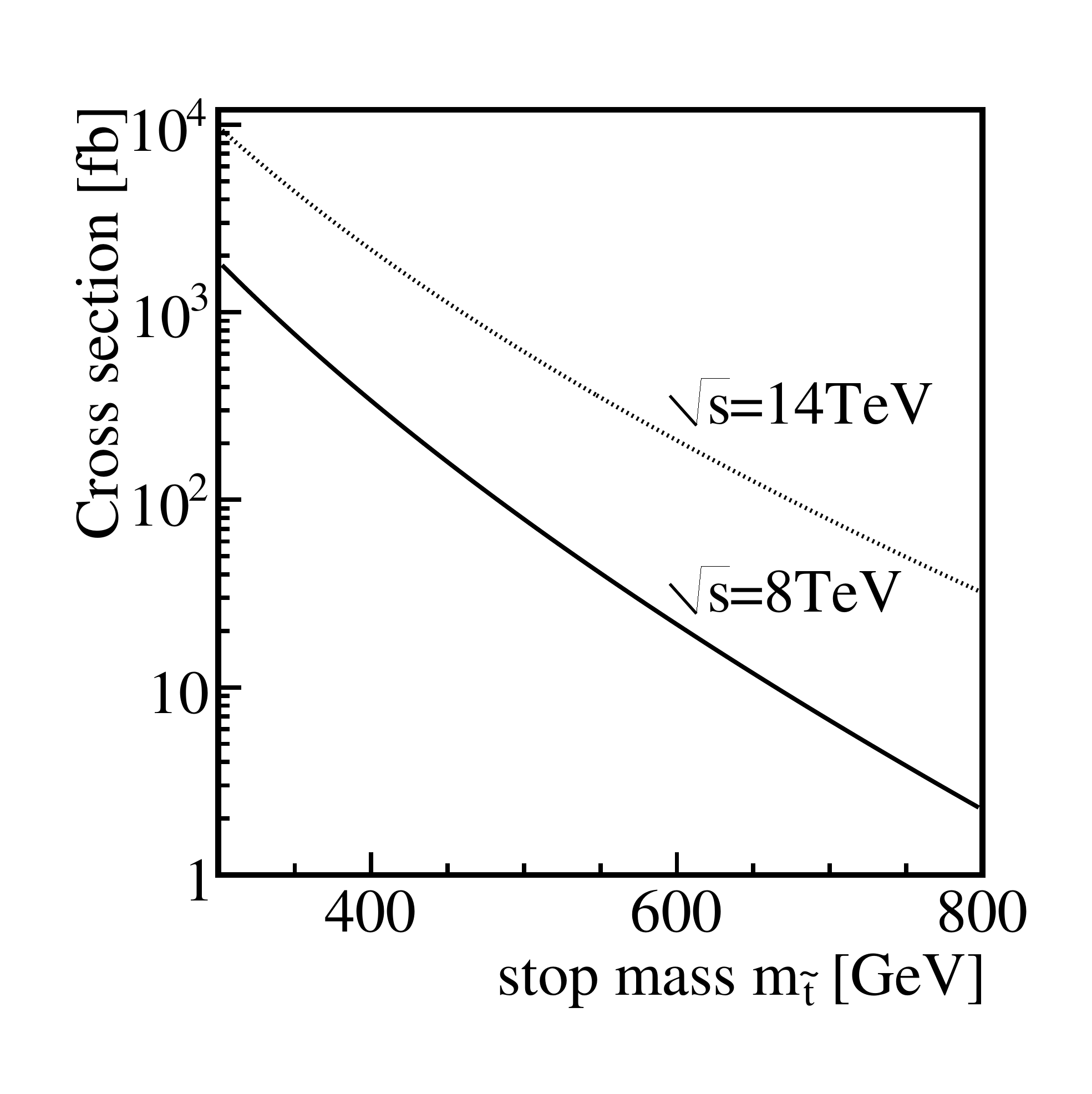}
\hfill
\raisebox{2cm}{
\begin{tabular}{l|rrrrrrr|r}
\hline
& \multicolumn{7}{c|}{$\st{} \st{}^*$} & \multicolumn{1}{c}{$t\bar{t}$} \cr
$\mst[\gev]$      & 300 &350 &400& 450 & 500 & 600 &700& \cr \hline
$\sigma_{14\tev}[\fb]$  & 9750 & 4380 &2150 & 1130 & 623 & 215 & 81.2 & 918000 \\
$\sigma_{8\tev}[\fb]$   & 1870 &760 & 337 &  160 & 80.5 & 23.0 & 7.19  & 234000 \\ \hline
\end{tabular} }
\vspace*{-7mm}
\caption{Production cross section for (s)top pairs as a function of
  the mass $\mst$ assuming $\sqrt{S}=14$ and $8$~TeV.}
\label{fig:cxn}
\end{figure}

Although the LHC production cross section for stop pairs is much
smaller at 8~TeV than at 14~TeV there is a fair chance to disentangle
this signal from the also significantly smaller SM backgrounds. For
the rest of the paper we omit the index `1' for the lighter of the two
stops. Stop pair production always refers to the pair production of
the lighter of the two stop mass eigenstates.  In Fig.~\ref{fig:cxn}
we show the NLO signal cross sections for $\sqrt{s}=8$ TeV and 14 TeV
as a function of the stop mass~\cite{stops_resummed}.  It drops by
roughly a factor 1/6 from 14~TeV to 8~TeV collider energy.  This
reduction is partly compensated for by the $t\bar{t}+$jets cross
section, which is reduced by a factor 1/4. We consistently normalize
our $t\bar{t}$ cross section to the approximated NNLO result of 234~pb
at 8~TeV and 918~pb at 14~TeV~\cite{tt_nnlo}.

In our 14~TeV analyses we have shown that relying on boosted hadronic tops
we can achieve $S/B \sim 1$ and $S/\sqrt{B} > 5$ in the fully hadronic
mode~\cite{HEP1} and $S/B \sim 2$ and $S/\sqrt{B} > 5$ in the
semi-leptonic mode~\cite{HEP2} for $10~\ifb$. For the 2012 run at
8~TeV the envisioned integrated luminosity also ranges around
$10-20~\ifb$. With the small production rates at 8~TeV we need to adapt
our search strategies to retain enough signal events to achieve a good
signal significance $S/\sqrt{B}$ for $10~\ifb$.\medskip

To decide on how to improve the analyses it is instructive to study
kinematic correlations between the two tops in the signal process
\begin{equation}
pp \to \st{} \st{}^*
   \to ( t \nz1) \, (\bar{t} \nz1)
\label{eq:signal}
\end{equation}
and the $t\bar{t}$ background.  Fig.~\ref{fig:cxn2d} shows the top
versus the anti-top transverse momenta for the signal and background,
assuming $\mst=400$~GeV. For the $t\bar{t}$ background we
find a strong correlation between the top and the anti-top, while for
tops from stop decays there is almost no correlation. The initially
back-to-back configuration of the stops is significantly distorted by
the $\nz1$ LSP momenta.

A sizable (transverse) top momentum is the key to top tagging.  In
Tab.~\ref{tab:cxn2d} we show the predicted rates for at least one top
with $p_{T,t}>200$~GeV. This is the value we need to reliably apply
subjet techniques and fully reconstruct hadronic tops~\cite{HEP3}. For
the top pair background we include up to two hard jets when we
simulate the distributions. We see that these additional jets lead to
a significant fraction of events where only one of the two tops drops
below 200~GeV. The signal with its comparably uncorrelated tops
clearly prefers only one strongly boosted top quark.

The last column in Tab.~\ref{tab:cxn2d} shows the ratios of rates for
400~GeV stops divided by the $t\bar{t}$ background rate.  Indeed, we
gain a factor 2.7 in $S/B$ by focusing on events with only one top
with $p_{T,t}>200$~GeV. This means that focusing on events with
asymmetric $p_{T,t}$ by only requiring one top tag should be the
strategy for the 2012 run at 8~TeV.  In the following subsections we
study four statistically independent samples or analyses:
\begin{itemize}
\item two hadronic boosted tops \\[-6mm]
\item one hadronic boosted top and one hadronic un-boosted top \\[-6mm]
\item one hadronic boosted top and one leptonic top \\[-6mm]
\item two leptonic tops
\end{itemize}
\medskip

\begin{figure}[t]
\includegraphics[height=0.30\textwidth]{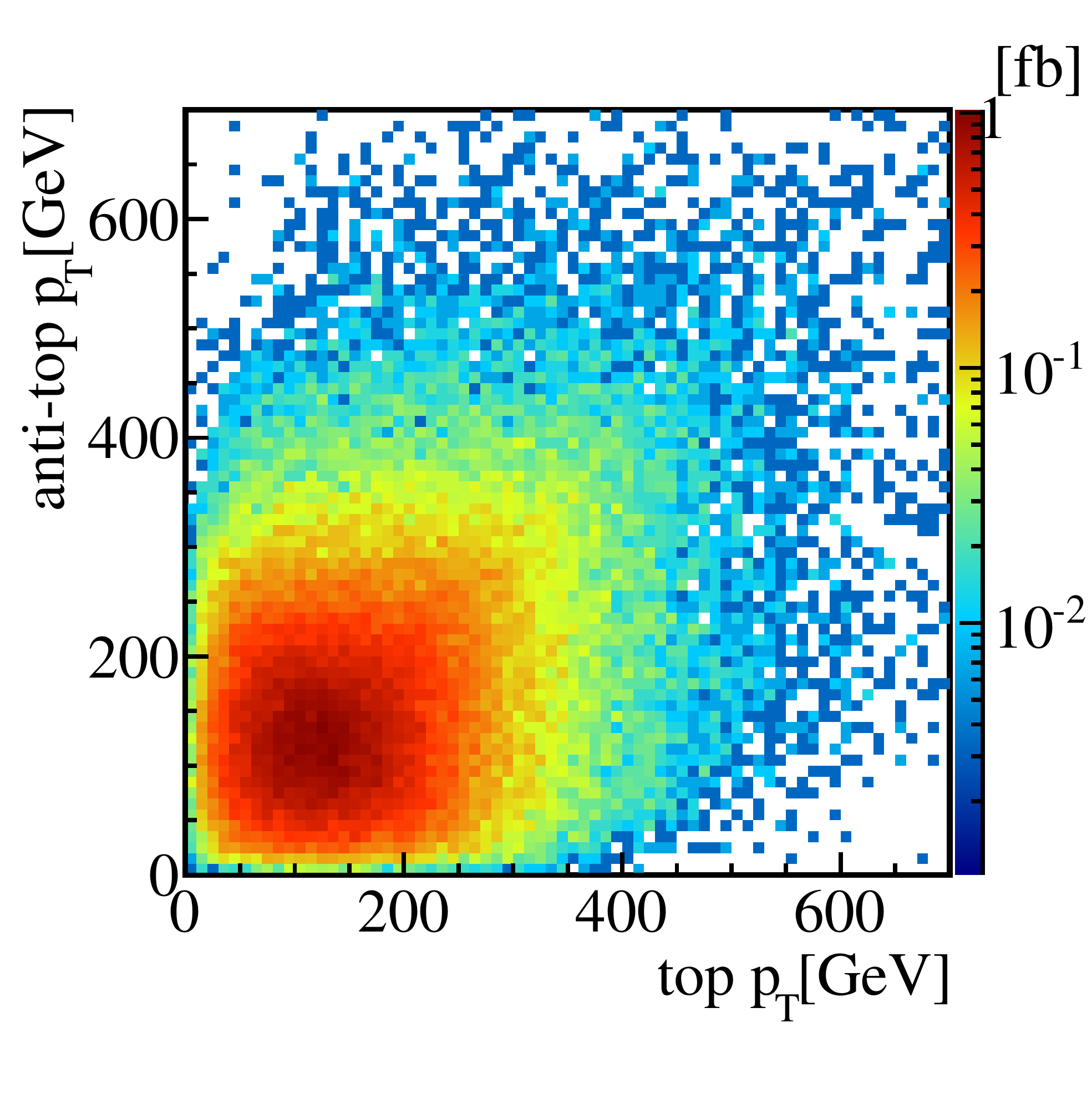}
\includegraphics[height=0.30\textwidth]{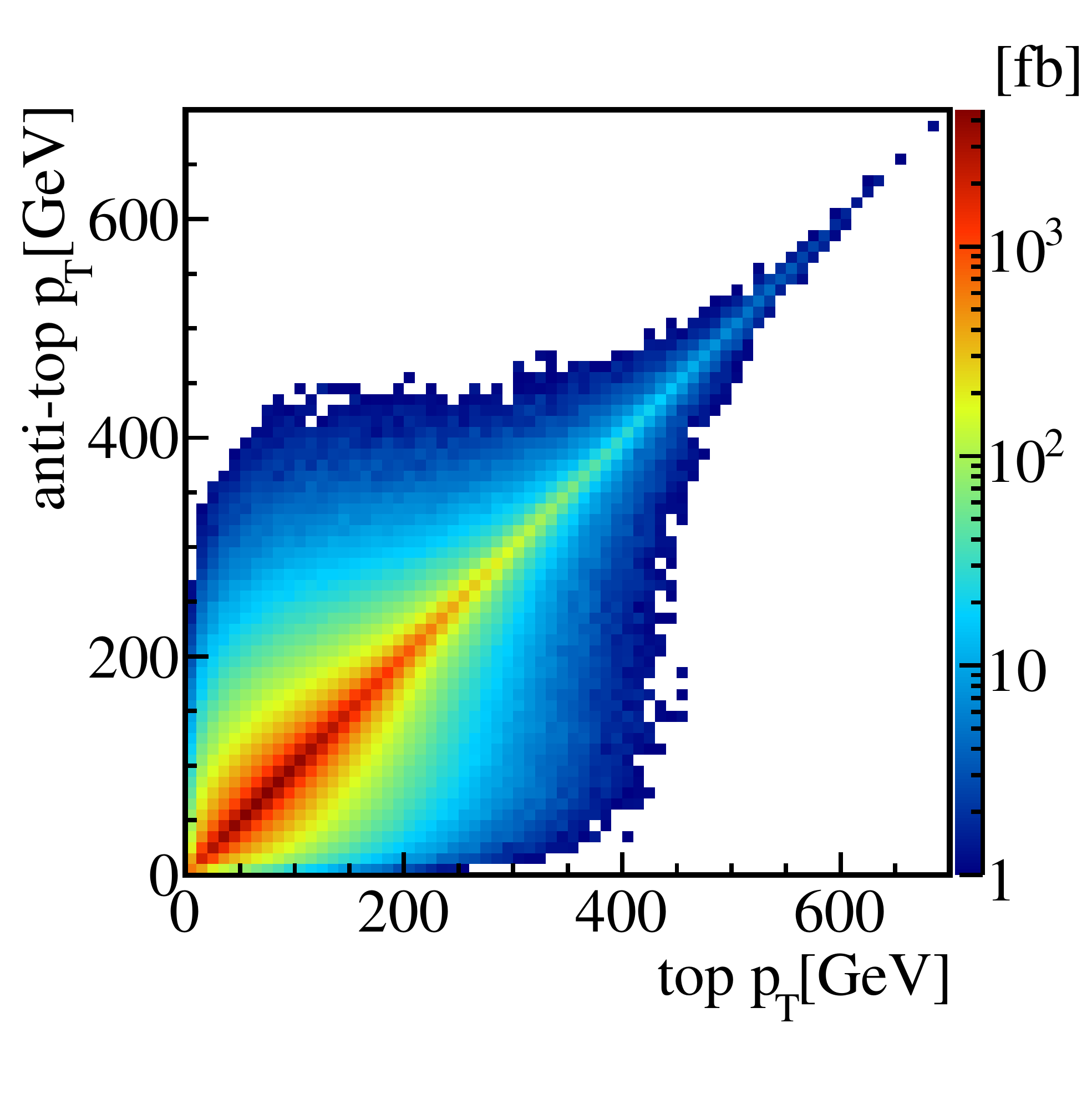}
\includegraphics[height=0.30\textwidth]{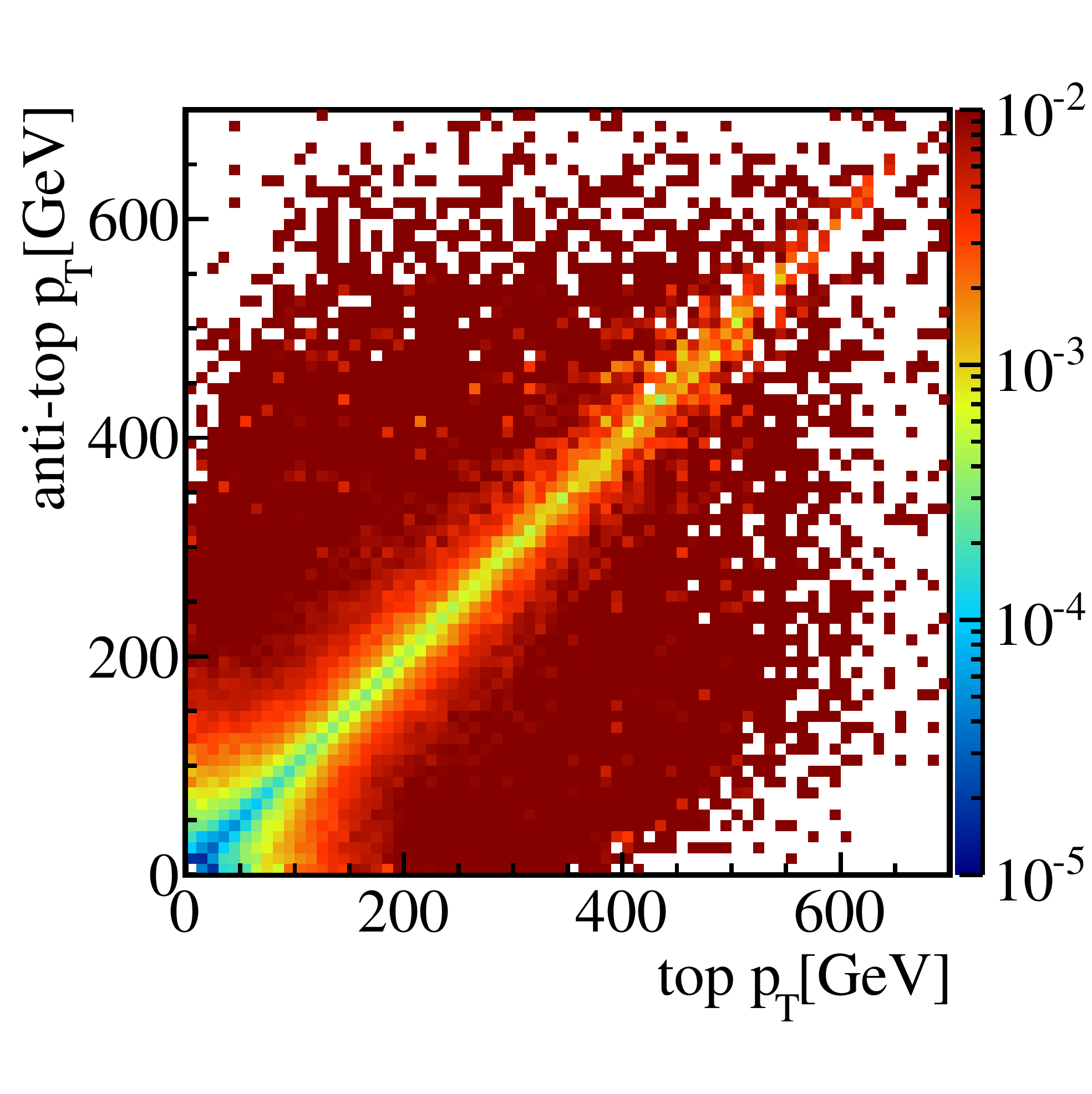}
\caption{Top vs. anti-top transverse momenta for 400~GeV stop pairs
  (left) and the $t\bar{t}$ background (center) at 8~TeV.  The right
  panel shows the ratio $\st{}\st{}^*/t\bar{t}$.}
\label{fig:cxn2d}
\end{figure}

\begin{table}[b]
\begin{tabular}{l|rrrrrr|r||c}
\hline 
$\sqrt{s}=8$~TeV  & \multicolumn{6}{c|}{$\st{} \st{}^*$}
& \multicolumn{1}{c||}{$t\bar{t}$} & $\sigma_{\st{}\st{}^*}/\sigma_{t\bar{t}}$  \\
\hline 
$\mst[\gev]$
& 350 & 400 & 450 & 500 & 600 & 700 &&\multicolumn{1}{c}{400} \\
\hline
at least one top with $p_{T,t}>200$ GeV 
&    252.21&    158.38&     96.83&     57.67&     19.80&    6.67&  $3.45 \cdot 10^4$ &  $4.6 \cdot 10^{-3}$   \cr \hline
only one top with $p_{T,t}>200$  GeV 
&    172.13&    109.63&     64.93&     36.77&     10.49&      2.80&  $1.57 \cdot 10^4$  & $7.0 \cdot 10^{-3}$ \cr
two tops with $p_{T,t}>200$  GeV 
& 80.07&     48.75&     31.90&     20.90&      9.30&      3.87&  $1.89 \cdot 10^4$  & $2.6 \cdot 10^{-3}$ \cr
\hline
\end{tabular}
\caption{
Signal and background cross sections [fb] for different stop masses. We assume BR($\st{}\to t \nz1) = 1$.}
\label{tab:cxn2d}
\end{table}

To generate the signal sample we rely on {\sc Herwig++}~\cite{herwig}
and assume stop masses of $\mst=350$, 400, 450, 500, 600, and
700~GeV. We normalize their production cross section to the {\sc
  Prospino} results at next-to-leading
order~\cite{prospino_stops}. They range from 0.8~pb ($\mst = 350$~GeV)
to 0.1~pb (500~GeV) as shown in Fig.~\ref{fig:cxn}. Resummation
slightly increases this rate further~\cite{stops_resummed}.  For all
stop masses we choose the same lightest neutralino mass
$\mne{1}=100$~GeV.\medskip

Our Standard Model backgrounds are $t\bar{t}$+jets, QCD jets,
$W$+jets, $Z$+jets, and $t\bar{t}Z$.  We use {\sc
  Alpgen+Pythia}~\cite{alpgen,pythia} to generate the corresponding
samples. For all processes except for $t\bar{t}Z$ production we use
MLM matching~\cite{mlm} to simulate additional hard radiation. We
match up to $t\bar{t}$+2~jets, $W$+3~jets, $Z$+4~jets, and $3 - 5$ jets
for the QCD sample.

The leading $t\bar{t}$+jets background sample we normalize to the
approximate NNLO rate of 234~pb~\cite{tt_nnlo}.  For the subleading
background channels we use the leading order normalization.  The
$t\bar{t}Z$ cross section at LO yields 21.5~fb, based on {\sc Alpgen}
including the $Z\to \nu\bar{\nu}$ branching ratio.  Since the
$t\bar{t}Z$ rate only becomes comparable to the stop rate for
$\mst \gtrsim 600$~GeV we neglect this irreducible $t\bar{t}Z$
background. We have checked that for all processes considered this
does not affect the quoted results.\medskip

Our analysis is based on a simple calorimeter simulation with
granularity of $0.1 \times 0.1$ in $(\eta, \phi)$.  We sum the four
momentum of all particles in each cell and rescale the resulting
three-momentum such as to make the cells massless. The calorimeter
cells are later on used as (fat)-jet constituents. Throughout this
work we use the Cambridge/Aachen (C/A) algorithm~\cite{ca_algo} with
$R=1.5$, as implemented in {\sc FastJet}~\cite{fastjet}. The resulting
fat jets are then used as input for the {\sc
  HEPTopTagger}. Preliminary ATLAS analysis show that the {\sc
  HEPTopTagger} results are only very mildly affected by detector
effects, underlying event, or pile-up~\cite{gregor}.  For regular QCD
jets use the same C/A algorithm with $R=0.5$.  When analyzing leptonic
or semileptonic top decays we require the leptons to be hard and
isolated: $p_{T,\ell} > 20$ GeV and $E_{T,\text{had}} < 0.1 \,
E_{T,\ell}$ within $R<0.2$ around the lepton.

\newpage

\section{Stop reconstruction at 8 TeV}
\label{sec:reconst}

As mentioned in the last section, we present four distinct signatures
of stop pairs decaying to a pair of top quarks plus missing
energy. The different analyses are chosen such that they are
statistically independent and can be combined, if required.

\subsection{Two top tags}
\label{sec:hadronic}

As a first attempt to apply the successful 14~TeV strategy to the 2012
run, we analyze fully hadronic stop pairs,
\begin{equation}
 pp \to \st{} \st{}^*
    \to ( t \nz1) \, (\bar{t} \nz1) 
    \to ( b j j \nz1) \, (\bar{b} j j \nz1) \; .
\end{equation}
One of the advantages of the fully hadronic mode is that, in
principle, the top tagger fully reconstructs both top momenta. This
means that the fully hadronic channel is particularly well suited for
detailed studies of a top partner signal.  For 14~TeV we indeed expect
signal-to-background ratios around $S/B \sim 1$~\cite{HEP1}. In this
section we follow the same analysis steps at 8~TeV.\medskip

\begin{table}[b]
{\small
\begin{tabular}{l|rrrrrr|rrrr||cc}
\hline 
$\sqrt{s}=8$~TeV & \multicolumn{6}{c|}{$\st{} \st{}^*$}
& $t\bar{t}$ 
& QCD
& $W$+jets 
& $Z$+jets 
& $S/B$ 
& $S/\sqrt{B}_{10\ifb}$ \\
\hline 
\hline
$\mst[\gev]$
& 350 & 400 & 450 & 500 & 600 & 700 &&&&&\multicolumn{2}{c}{400} \\
cross section [fb]
&   760 &    337 &    160 &     80.5&     23.0&      7.19
& $2.3 \cdot 10^5$ & $6.5 \cdot 10^8$ & $1.6 \cdot 10^6$ &  $1.2\cdot 10^4$ &$< 10^{-6}$ & $0.04$ \\
\hline
$\ell$ veto 
&    488&    215&    101&     50.5&     14.4&      4.46
&$1.6 \cdot 10^5$& $6.5 \cdot 10^8$&$1.3 \cdot 10^6$&  $1.2\cdot 10^4$ & $< 10^{-6}$&0.03\cr
\hline
$n_\text{fat} \ge 2$ 
&    167&     88.3&     48.0&     26.6&      8.71&      2.96
&  $3.7 \cdot 10^4$ & $2.0 \cdot 10^{7}$ & $1.1 \cdot 10^5$& $1.3\cdot 10^3$  & $<10^{-5}$ & 0.06 \cr
$\ptmiss>100$ GeV 
& 104&     65.0&     38.5&     22.5&      7.76&      2.74
&   $1.6 \cdot 10^3$ & $2.0 \cdot 10^{5}$&   $1.9\cdot 10^3$&    694 & $3 \cdot 10^{-4}$ & 0.45    \cr
$ n_\text{tag} \ge 1 $
&27.5&     18.5&     11.87&      7.60&      2.91&      1.12&    375 & $2.5\cdot 10^3$&     36.7&     17.0 &  $6 \cdot 10^{-3}$ & 1.1  \cr
$  n_\text{tag} \ge 2$
&      2.34&      1.65&      1.12&      0.76&      0.34&      0.14&      6.40&   18&      0.5&      -- & 0.07 & 1.0 \cr
$b$-tag inside top 
&      0.74&      0.58&      0.35&      0.25&      0.11&      0.05&      1.93&     0.18&      -- &      --  & 0.27 &1.3  \cr
$  m_{T2}>250$ GeV       
&      0.24&      0.30&      0.22&      0.18&      0.09&      0.04&      0.34&       0.03 &      -- &      --  & 0.79 & 1.5 \cr
\hline
\end{tabular}
}
\caption{Analysis flow for the two-top analysis.  All numbers are
  given in fb. The symbol ``--'' denotes less than $0.01~$~fb.}
\label{tab:2boosted}
\end{table}

For the fully hadronic double tag mode we veto isolated leptons and require at
least two fat jets with
\begin{equation}
p_{T,j} > 150/150~\gev \qqquad \text{and} \qqquad \met > 100~\gev \;,
\label{eq:cuts1}
\end{equation}
where both numbers are slightly reduced compared to the 14~TeV
case~\cite{HEP1}.

In Tab.~\ref{tab:2boosted} we show the numbers of events after each
cut. For illustration purposes we compute $S/\sqrt{B}$ for $10~\ifb$
after each step, taking into account only statistical errors. 

We first require missing transverse momentum, just like for almost all
supersymmetry analyses.  To simulate the amount of fake missing energy
in the pure QCD jets background we would need a detailed detector
simulation. Instead, we adopt a conservative efficiency of 1\% for QCD
events to pass this cut. With this assumption QCD jets are our
dominant background, which essentially forces us to apply a $b$-tag
inside the first tagged top later on.

This analysis is based on two top tags, using the {\sc HEPTopTagger}
algorithm. Already at this second stage the $W$+jets and $Z$+jets
backgrounds become negligible and we are left with $t\bar{t}$ and QCD
jets. On the other hand, all rates are already down to a femto-barn
level.  As mentioned above, we need an additional tagged $b$-subjet
inside the first tagged top, to reduces the QCD background to a
negligible level. It has been shown that $b$-tagging is not a problem
inside a fat jet, so we apply a $b$-tagging efficiency of 50\%
for $b$-quark and a 1\% fake rate for light quarks and
gluons~\cite{giacinto}. In the final step the $t\bar{t}$ background
can be reduced by requiring
\begin{equation}
  m_{T2}>250~\gev \; . 
\label{eq:cuts2}
\end{equation}
This way we achieve $S/B=0.79$ and $S/\sqrt{B}=1.5$ for an integrated
luminosity of $10~\ifb$. The fact that the significance is reduced by
a factor 1/3 as compared to the 14~TeV run is explained by the factor
1/6 in signal cross sections and the softer $p_{T,t}$ spectrum.\medskip

To study ways out of this rate limitation we can turn to the top
tagging efficiencies.  After requiring $\met > 100$~GeV we show these
efficiencies in Tab.~\ref{tab:2boosted}. For the first (mis-)tag it
ranges around 30\% for the signal, 20\% for $t\bar{t}$, and 1\% to 2\%
for QCD jets, $W$+jets, and $Z$+jets.  For the second tag the
efficiencies are around 10\% for the signal, 2\% for $t\bar{t}$, while
QCD jets, $W$+jets, and $Z$+jets remain at 1 to 2\%. The reason for
this small $t\bar{t}$ efficiency is that in events with large missing
momentum one of the tops has to decay leptonically. The second tag
then is a fake-top from the remaining $b$ jet combined with hard QCD
radiation. Hence the second top tag is very helpful against the leading top pair background. Unfortunately due to the already small signal rate, the second top tag does not increase the sensitivity significantly.

\subsection{One top tag and one bottom tag}
\label{sec:oneboost}

\begin{table}[b]
{\small
\begin{tabular}{l|rrrrrr|rrrr||cc}
\hline 
$\sqrt{s}=8$~TeV & \multicolumn{6}{c|}{$\st{} \st{}^*$}
& $t\bar{t}$ 
& QCD
& $W$+jets 
& $Z$+jets 
& $S/B$ 
& $S/\sqrt{B}_{10\ifb}$ \\
\hline 
\hline
$\mst[\gev]$
& 350 & 400 & 450 & 500 & 600 & 700 &&&&&\multicolumn{2}{c}{400} \\
\hline
$\ell$ veto, $n_\text{fat} \ge 1$
&    378&    186&     92.3&     47.8&     14.0&      4.40&  $6.9 \cdot 10^4$& $3.8 \cdot 10^7$ & $1.9 \cdot 10^5$&  $5.0 \cdot 10^4$ & $5\cdot 10^{-6}$ & 0.1\cr
$\ptmiss>100$ GeV
&    264&    149&     78.6&     42.1&     12.9&      4.15&  $7.1 \cdot 10^3$&$3.8 \cdot 10^5$ &  $1.3 \cdot 10^4$ &   $3.2 \cdot 10^3$ & $4\cdot10^{-4}$ & 0.7\cr
$ n_\text{tag} \ge 1$
&     48.8&     32.6&     19.9&     12.0&      4.29&      1.54&    959& $2.7\cdot 10^3$  &    106&     57.3 &$9\cdot 10^{-3}$&1.7\cr
$n_\text{tag}=1$, $b$-tag inside 
&     13.0&      8.57&      5.34&      3.14&      1.15&      0.42&    322&  26.4  &     1.05 &     0.57 &0.024 &1.4\cr
additional $b$-tag
&     4.41&      2.81&      1.75&      1.04&      0.39&      0.15&    116&    0.26&      0.01 &      -- &0.024& 0.82 \cr
$ m_T^{(b)}>200$ GeV 
&      0.92&      0.90&      0.73&      0.50&      0.24&      0.10&      1.20&      -- &       -- &      --  &0.73 & 2.6 \cr \hline
($\tau$ rejection)
&      0.89&      0.89&      0.71&      0.49&      0.23&      0.10&      0.85&    -- &      -- &      -- & 1.00 &  3.0 \cr
\hline
\end{tabular}
}
\caption{Analysis flow for one top tag and one $b$-tag.  All numbers
  are given in fb. The symbol ``--'' denotes less than $0.01~$~fb. In
  the last line we illustrate the potential of a 100\% efficient tau
  veto.}
\label{tab:boostedhad_b}
\end{table}

To improve the fully hadronic analysis presented in the last section
we propose a search for one boosted top and one $b$-tag in the
recoiling softer top decay jets. As a starting point we apply a lepton
veto and require exactly one fat jet together with missing transverse
momentum,
\begin{equation}
p_{T,j} > 150~\gev \qqquad \text{and} \qqquad \met > 100~\gev \;.
\label{eq:tbcuts1}
\end{equation}
One subjet inside the tagged top has to be $b$-tagged. In addition, we
require a continuum $b$-tag which cannot be a constituent of the
tagged top.  In Tab.~\ref{tab:boostedhad_b} we see that after these
two $b$-tags all backgrounds except for top pair production are
negligible.\medskip

To reduce the still overwhelming $t\bar{t}$ background we construct a
specific transverse mass variable from the general form
\begin{equation}
m_T(p_\text{vis}, \ptmiss)= \sqrt{m^2_\text{vis} + 2 |\ptmiss| (E_{T,\text{vis}} - p_{T,\text{vis}} \cos \phi)} \; ,
\label{eq:mt}
\end{equation}
where $\phi$ is the angle between the transverse visible and missing
momenta.  In events with an isolated lepton and missing momentum this
variable is commonly used to reject leptonic $W$ decays because $m_T$
is bounded from above.

\begin{figure}[t]
\includegraphics[width=0.40\textwidth]{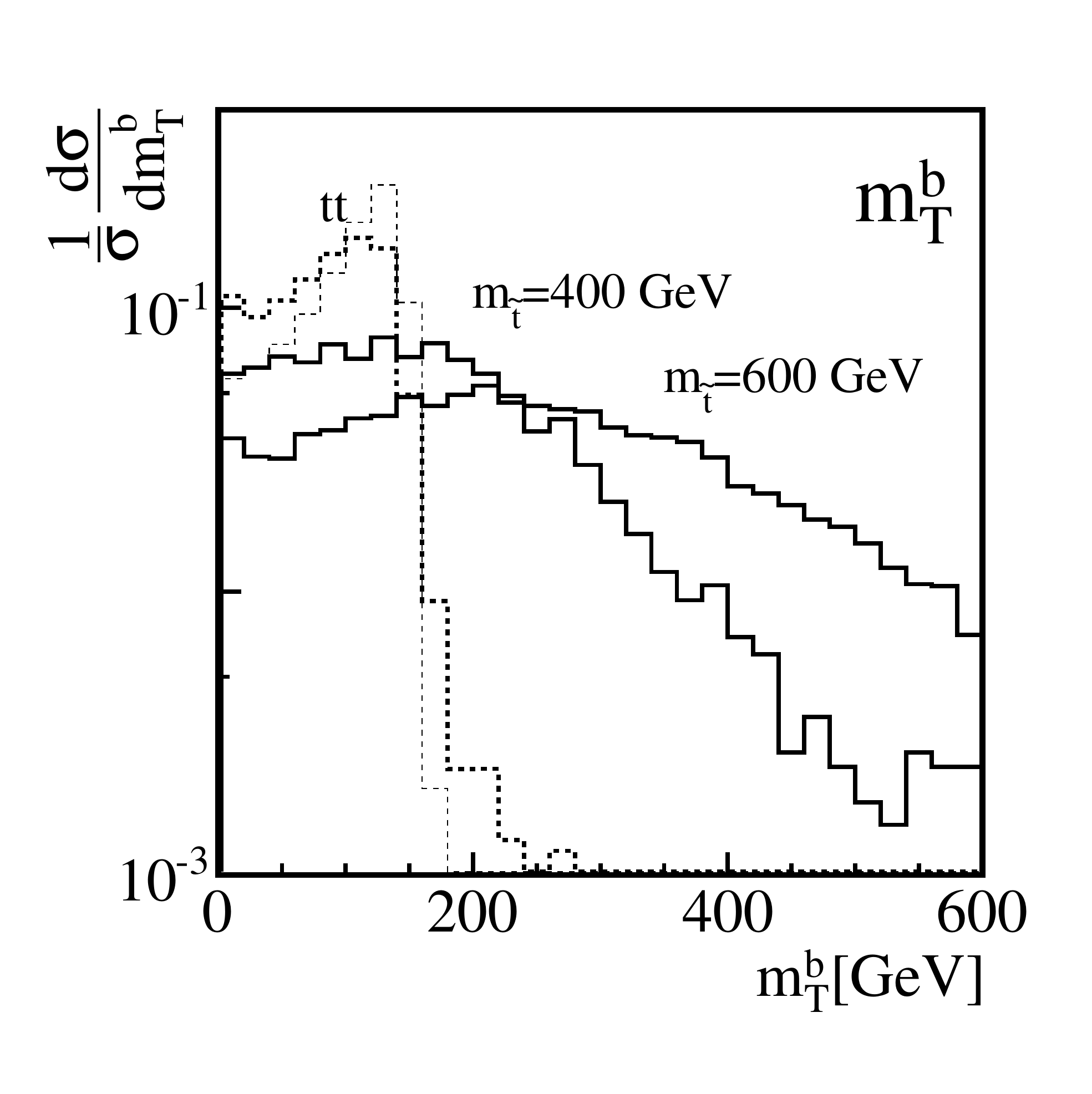}
\hspace*{0.1\textwidth}
\includegraphics[height=0.40\textwidth]{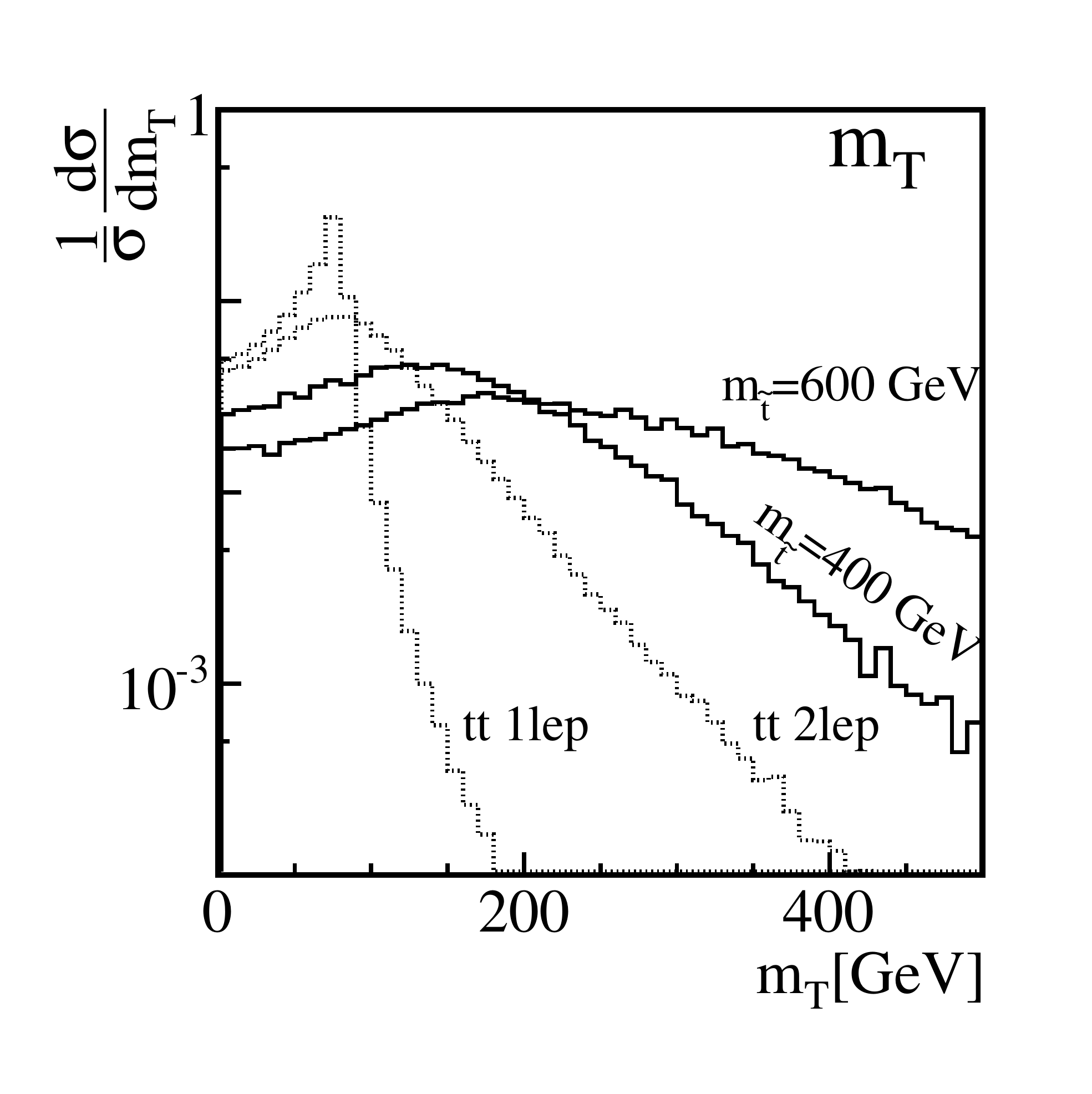}
\vspace*{-6mm}
\caption{Left: $m_T^{(b)}$ distribution for the hadronic signal
  (solid) and the $t\bar{t}$ background (dashed).  The thin dashed
  line shows $t\bar{t}$ semi-leptonic events at parton level. Right:
  $m_T$ distribution for the signal, for the semi-leptonic top pairs and for di-leptonic top
  pairs. The distributions are based on events which have exactly one
  lepton $(e,\mu)$ in the final state.}
\label{fig:mt}
\end{figure}
 
The main background in the no-lepton mode with large missing momentum
comes from $t\bar{t}$ events where one top decays through a tau
lepton. In these events large missing momentum can be induced by the
neutrinos from the $W$ and $\tau$ decays. To separate these events
from the signal with its two neutralinos we construct the transverse
mass with a $b$-jet instead of the lepton, and require
\begin{equation}
m_T^{(b)}  \equiv m_T(p_b, \ptmiss) > 200~\gev.
\label{eq:tbcuts3}
\end{equation}
The left panel of Fig.~\ref{fig:mt} shows the normalized $m_T^{(b)}$
distributions for the signal and the $t\bar{t}$ background.  As
expected, there is an endpoint at $m_t$ for the background.  As a cross
check the thin dashed line shows the $m_T^{(b)}$ distribution from the
missing momentum and the bottom momentum from a leptonic top decay in
semi-leptonic $t\bar{t}$ events.  This parton level distribution shows
good agreement with the background distribution from the hadronic
final state.  We also tested a similar $m_{T2}^{(b)}$, but the remaining
number of signal events turns out to be too small.\medskip

After this final $m_T^{(b)}$ cut all signal and background rates shown
in Tab.~\ref{tab:boostedhad_b} are again at the fb level.  For a
400~GeV stop we find $S/B=0.73$ and $S/\sqrt{B}=2.6$ with $10~\ifb$ of
data.  Because we know the origin of the remaining $t\bar{t}$ events
we could further improve the results by rejecting tau leptons. Just to
illustrate the potential of such a requirement, we assume a $100 \%$
efficiency for tau rejection in the last line of
Tab.~\ref{tab:boostedhad_b}.

\subsection{One top tag and one lepton}
\label{sec:leptonic}

\begin{table}[b]
{\small
\begin{tabular}{l|rrrrrr|rrr||cc}
\hline 
$\sqrt{s}=8$~TeV & \multicolumn{6}{c|}{$\st{} \st{}^*$}
& $t\bar{t}$ 
&$t\bar{t}Z$
& $W$+jets 
& $S/B$ 
& $S/\sqrt{B}_{10\ifb}$ \\
\hline 
\hline
$\mst[\gev]$
& 350 & 400 & 450 & 500 & 600 & 700 &&&&\multicolumn{2}{c}{400} \\
cross section [fb]
&  760&    337&    160&     80.5&     23.0&      7.19 & $2.3 \cdot 10^5$ &21.5 & $1.6 \cdot 10^6$  &\\
\hline
$n_\ell =1$
  & 241&    108&     52.3&     26.5&      7.58&      2.39&  $6.9 \cdot 10^4$ &  6.24&  $2.8 \cdot 10^5$ \cr
$n_\text{fat} \ge 1$ 
&    145&     76.5&     40.6&     22.1&      6.83&      2.24&  $2.4  \cdot 10^4$ & 3.21 & $3.7 \cdot 10^4$ \cr
$\ptmiss>100$ GeV 
&    104&     61.5&     34.8&     19.5&      6.28&      2.11&   5631& 2.20&      8547 \cr
$ n_\text{tag}=1$
& 13.1&      9.02&      5.80&      3.60&      1.33&      0.50&    789&  0.33&     80.5&  0.01 & 1.0   \cr
$  m_T>150$ GeV &      4.63&      4.27&      3.25&      2.19&      0.94&      0.38&       3.28 &   0.10&        0.99&   1.0  & 6.5 \cr
\hline
$b$-tag inside top  &      1.47 &      1.38&      1.06 &      0.70&      0.31&      0.13&     0.63 & 0.03 & -- &  2.1 & 5.4 \cr
\hline
\end{tabular}
}
\caption{Analysis flow for one top tag and one lepton.  All numbers
  given in fb.  The symbol ``--'' denotes less than $0.01$~fb.}
\label{tab:semilep}
\end{table}

Following the previous section, events with one boosted and one
non-boosted top are well suited to extract stop pairs from Standard
Model backgrounds. An obvious question then becomes what happens if
the softer of the two tops decays leptonically,
\begin{equation}
 pp \to \st{} \st{}^*
    \to ( t \nz1) \, (\bar{t} \nz1) 
    \to ( b \ell^+ \nu \nz1) \, (\bar{b} j j \nz1) 
       +( b j j \nz1) \, (\bar{b} \ell^- \bar{\nu} \nz1).
\end{equation}
This time we require one isolated lepton, a sizable amount of missing
energy and one fat jet with a top tag. In Table~\ref{tab:semilep} we
see that the $t\bar{t}$ background is still overwhelming. The reason
is a significant fraction of semi-leptonic $t\bar{t}$ events passing
these cuts. 

The transverse mass, Eq.\eqref{eq:mt}, has an upper kinematic endpoint
for events where the missing energy comes from leptonic $W$ decays.
To efficiently reject leptonic top pair events as well as any kind of
$W$ events we require~\cite{Aad:2011wc}
\begin{equation}
m_T > 150~\gev \; .
\end{equation}
Surprisingly, a non-negligible number of $t\bar{t}$ events survives
this cut.  Similarly to the last section, purely leptonic top pairs
can fake a top tag from the $b$ jets and additional QCD radiation.
The right panel of Fig.~\ref{fig:mt} shows the normalized $m_T$
distributions for the signal, for semi-leptonic $t\bar{t}$ events, and
for purely leptonic $t\bar{t}$ events.

After imposing all the above cuts we arrive at a promising
signal-to-background ratio of $S/B=1$ and a significance of
$S/\sqrt{B}=6.5$ for $10~\ifb$ of 8~TeV running. Unlike the hadronic
channels this analysis does not require a $b$ inside or outside the
top tag. However, if we are willing to pay the price in available rate
we can apply the usual $b$-tag among the top tag constituents.

\subsection{Two leptons}

Until now, all our stop pair analyses involve one boosted hadronic top
decay which we identify and reconstruct using a top tagger. If we
loosen our requirements on event reconstruction we can of course
search for top pairs in purely leptonic top pairs,
\begin{equation}
 pp \to \st{} \st{}^*
    \to ( t \nz1) \, (\bar{t} \nz1) 
    \to ( b \ell^+ \nu \nz1) \, ( \bar{b} \ell^- \nu \nz1) \; . 
\end{equation}
This di-lepton channel turns out to have the largest reach in finding
or ruling out anomalies in the top sector. On the other hand, in the
absence of any reconstructed mass a deviation from the Standard Model
cannot confirm the existence of top partners. Therefore, we consider
the di-lepton channel a very powerful tool to confirm and
statistically support any anomaly found in one of the hadronic
channels.\medskip

\begin{figure}[t]
\includegraphics[height=0.40\textwidth]{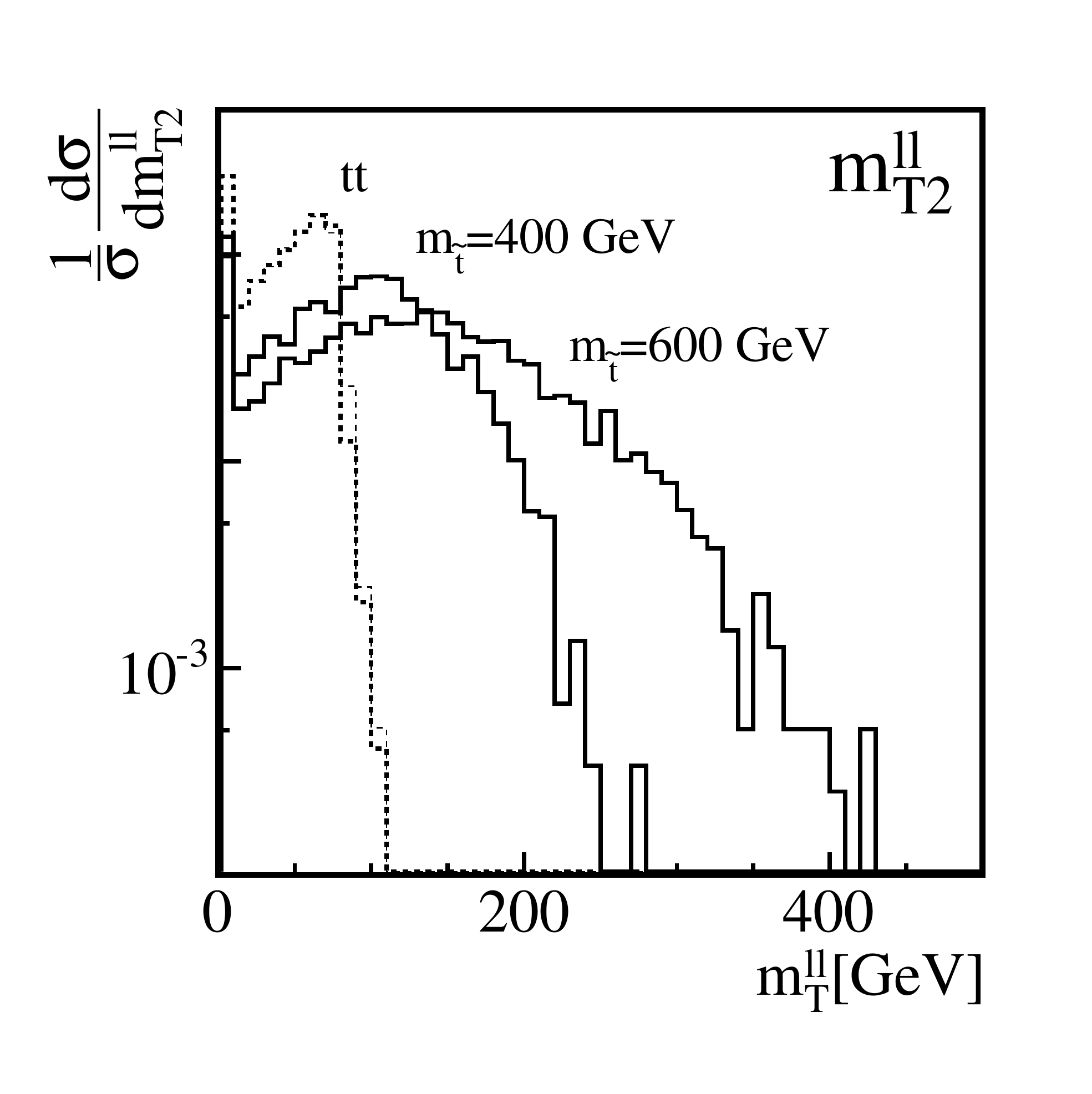}
\vspace*{-7mm}
\caption{Normalized $m_{T2}^{\ell\ell}$ distributions for the stop
  signal and the $t\bar{t}$ background. The light dotted line for the
  $t\bar{t}$ background includes a Gaussian smearing of the missing
  energy.}
\label{fig:mt2_dilep}
\end{figure}

\begin{table}[b]
\begin{tabular}{l|rrrrrr|ll||cc}
\hline 
$\sqrt{s}=8$~TeV & \multicolumn{6}{c|}{$\st{} \st{}^*$} 
& $t\bar{t}$ &  $t\bar{t}Z$ & $S/B$  & $S/\sqrt{B}_{10\ifb}$ \\
$\mst[\gev]$ & 350 & 400 & 450 & 500 & 600 & 700 &&&\multicolumn{2}{c}{400} \\
\hline
\hline
$n_{\ell}=2$
&     31.0&     14.3&      7.07&      3.58&      1.04&      0.33&   7651&      n.a. \cr 
$\ptmiss>100$GeV 
&     19.0&      9.99&      5.40&      2.94&      0.91&      0.30&   1313&      0.35\cr
\hline
$m_{T2}^{\ell \ell}>100$ GeV &      6.05&      4.30&      2.70&      1.65&      0.56&      0.20&      0.65 (0.79)&      0.09 & 5.8 (4.9) & 15.8 (14.5) \cr
$m_{T2}^{\ell \ell}>150$ GeV &      0.81&      1.21&      1.06&      0.81&      0.34&      0.14&      0.00 (0.03)&      0.02 & n.a. & n.a. \cr
\hline
\end{tabular}
\caption{ Analysis flow for the di-lepton mode. All numbers are given
  in fb.  The $t\bar{t}Z$ numbers are shown including the decay $Z \to
  \nu\bar{\nu}$. The number in parentheses include a smeared
  transverse momentum measurement.}
\label{tab:dilepmt2}
\end{table}

The previous sections show that the transverse mass $m_T$ with a
lepton or a $b$-jet momentum efficiently reduces semi-leptonic
$t\bar{t}$ backgrounds. For events with two sources of missing energy
a better-suited variable is~\cite{mt2}
\begin{equation}
m_{T2}^{\ell \ell}=\min_{\ptmiss=\pmiss_1 + \pmiss_2}
\left[ \max \left( m_T(p_{\ell_1}, \pmiss_1), m_T(p_{\ell_2}, \pmiss_2) \right)
\right] \; .
\label{eq:tbcuts2}
\end{equation}
In this definition the two transverse mass values, Eq.\eqref{eq:mt},
are computed with one of the two observed leptons. The missing momenta
in the two transverse mass values are a hypothetical split of the
observed missing transverse momentum into two parts. The mass of the
unobserved particles we assume to vanish. 

To enhance the di-lepton mode we need to identify additional sources
of missing energy, as compared to $t\bar{t}$ or
$W^+W^-$ events with their $W$ decay neutrinos. The particles
contributing to $m_{T2}$ from $t\bar{t}$ events come from the $W^+W^-$
subsystem, so for top quarks close to threshold the upper endpoint is
$m_{T2} < m_W$~\cite{Cohen:2010wv,Kats:2011qh}. In contrast, for the stop signal such an endpoint
does not exist. The corresponding $m_{T2}$ distributions we show in
Fig.~\ref{fig:mt2_dilep}. The leading detector effect is the smeared
missing momentum according to a Gaussian with $\sigma({\pmiss}) =
a\cdot \sqrt{\sum E_T}, a=0.53 \sim 0.57$~\cite{Aad:2009wy}.  The thin
dotted line in Fig.~\ref{fig:mt2_dilep} includes this smearing and
shows a slightly enhanced tail.\medskip

Our purely leptonic event selection starts with two isolated leptons
and $\ptmiss > 100$~GeV. Tab.~\ref{tab:dilepmt2} shows that the
$t\bar{t}$ background becomes essentially negligible after we require
\begin{equation}
m_{T2}^{\ell\ell}>100~\gev.
\end{equation}
Similarly, the total $t\bar{t}Z$ rate is significantly smaller than
the signal. Including a smeared missing energy measurement increased
the $t\bar{t}$ background from 0.65 to 0.79 fb.  For $\mst=400$ GeV,
we expect more than 40 signal events, giving $S/B =5.8$ and
$S/\sqrt{B} =15.8$ with $10~\ifb$.  A slightly harder cut $m_{T2}^{\ell
  \ell}>150$~GeV removes essentially all the SM backgrounds.  What is
most impressive is that this analysis completely ignores any jet which
might come with the $\ell^+ \ell^- \met$ system!

\section{Summary}
\label{sec:conclusion}

Reasonably light top partners are necessary to solve the hierarchy
problem. Therefore, searches for stops or other top partners are of
paramount interest to LHC physics. In 2012 the LHC will gather at
least $\mathcal{O}(10)~\ifb$ of data at 8~TeV. For four independent
search channels we show how 2012 data will start to either find or
exclude light top partners, decaying to top quarks and missing
energy.\medskip

In the fully hadronic mode we study two strategies: tagging either
one or two hadronic tops we find $S/B\sim1$ for a stop mass of
400~GeV. Unfortunately, the statistical significance is rather modest,
$S/\sqrt{B}=1.5$ (two tags) and $S/\sqrt{B}=3.0$ (one tag).

Searches for semi-leptonic or fully-leptonic top pairs are more
promising.  In the semi-leptonic mode we tag one top recoiling against
an isolated lepton. After cutting on $m_T$ we find $S/B=2.1$ and
$S/\sqrt{B}=5.4$.  In the di-lepton mode a cut on $m_{T2}^{\ell \ell}$
rejects almost all Standard Model backgrounds. This gives us a
striking sensitivity of $S/B=5.8$ and $S/\sqrt{B}=15.8$.

Obviously, the fully leptonic mode is unlikely to conclusively
reconstruct and confirm a top partner. However, the combination
with the statistically less significant hadronic modes should allow us
to establish a top partner signal in 2012.


\end{document}